\documentclass[reprint, amsmath,amssymb, aps, superscriptaddress]{revtex4-1} 
\usepackage{graphicx}
\usepackage{dcolumn}
\usepackage{bm}
\usepackage{color}
\usepackage[dvipdfmx, hypertex]{hyperref}

\begin{document}
\preprint{IPMU14-0187}

\title{Bootstrapping phase transitions in QCD and frustrated spin systems}


\author{Yu Nakayama}
\author{Tomoki Ohtsuki}
\affiliation{Kavli Institute for the Physics and Mathematics of the Universe (WPI),  \\ Todai Institutes for Advanced Study,
University of Tokyo, \\ 
5-1-5 Kashiwanoha, Kashiwa, Chiba 277-8583, Japan} 

\date{\today}

\begin{abstract} 
In view of its physical importance in predicting the order of chiral phase transitions in  QCD and frustrated spin systems, we perform the conformal bootstrap program of $O(n)\times O(2)$-symmetric conformal field theories in $d=3$ dimensions with a special focus on $n=3$ and $4$. The existence of renormalization group fixed points with these symmetries has been controversial over years, but our conformal bootstrap program provides the non-perturbative evidence. In both $n=3$ and $4$ cases, we find singular behaviors in the bounds of scaling dimensions of operators in two different sectors, which we claim correspond to chiral and collinear fixed points, respectively. In contrast to the cases with larger values of $n$, we find no evidence for the anti-chiral fixed point.  Our results indicate the possibility that the chiral phase transitions in QCD and frustrated spin systems are continuous with the critical exponents that we predict from the conformal bootstrap program.
\end{abstract}
\pacs{Valid PACS appear here}
\maketitle 
\section{Introduction} 
One of the greatest challenges to our human intellect is to understand the universal nature of phase transitions and critical phenomena. 
A significant triumph of the last century in this direction is the introduction of renormalization group (RG), which enables us to explain the infrared behaviors of physical systems from a small number of relevant parameters (see e.g. \cite{Cardy:1996xt}\cite{Pelissetto:2000ek}). If a system undergoes a continuous phase transition, its critical behavior is governed by the endpoint of the RG flow realized as a scale invariant and, in most cases, conformal invariant field theory. Whether such a conformal field theory (CFT) with a given symmetry of the system exists or not is of theoretical interest because without its existence the continuous phase transition cannot occur.

From this perspective, (non-)existence of fixed points in  $O(n)\times O(m)$-symmetric Landau-Ginzburg-Wilson (LGW) Hamiltonian in $d=3$ spatial dimensions
\begin{align}
\mathcal{H} &=   (\partial_\mu \phi^\alpha_a)(\partial_\mu \phi^\alpha_a) \cr
&+ u (\phi^{\alpha}_a\phi^\alpha_a)^2 + v(\phi^\alpha_a\phi^\alpha_b \phi^\beta_a \phi^\beta_b  - \phi^\alpha_a\phi^\alpha_a \phi^\beta_b \phi^\beta_b) , \label{Hamiltonian}
\end{align}
where $a=1,\cdots n$ and $\alpha = 1 \cdots m$,
 plays a crucial role in some physically important systems.  As a notable example, it has been argued that a $n$-component spin system placed on two-dimensional triangular lattices piled up into the vertical direction possesses $O(n) \times O(2)$ symmetry with the effective Hamiltonian \eqref{Hamiltonian} with $v>0$ (see \cite{Kawamura} for a review and earlier discussions on RG studies), and there exist rich experimental realizations for $n=2$ and $3$  (see \cite{Delamotte:2003dw}\cite{delamotte2005fustrated} for a convenient summary). Furthermore, in two-flavor QCD with the axial anomaly suppressed, the chiral phase transition is described by \eqref{Hamiltonian} with $v<0$ with $O(4) \times O(2) \simeq  SU(2)_L \times SU(2)_R \times U(1)_A$ symmetry. The possibility for the effective restoration of $U(1)_A$ symmetry at the chiral phase transition temperature has attracted a renewed interest. See e.g. \cite{Bazavov:2012qja}\cite{Aoki:2012yj}\cite{Cossu:2013uua}\cite{Buchoff:2013nra}\cite{Pelissetto:2013hqa}\cite{Ishikawa:2013tua} and reference therein for recent discussions.

Despite their importance, theoretical studies of LGW models \eqref{Hamiltonian} have been notoriously hard and there remain long-standing controversies regarding what kind of fixed points actually exist.
In \cite{pelissetto2001critical}, they derived the perturbative series directly at the physical dimension $d=3$ in massive-zero-momentum (MZM) scheme up to six-loop order, and in \cite{Calabrese:2003ww} minimal subtraction ($\overline{\mathrm{MS}}$) scheme up to five-loop order.  After careful resummations, the presence of non-trivial fixed points distinct from the Heisenberg ones $(v=0)$ were predicted for $n=2,3,4$ in \cite{pelissetto2001critical}\cite{Calabrese:2002af}\cite{Calabrese:2004nt}\cite{Pelissetto:2013hqa}, but some of the results have been criticized e.g. in \cite{Delamotte:2010ba} due to the lack of confirmation in the weak-coupling regime and the parameter dependence in the resummation. On the other hand, the functional RG truncated beyond the local potential approximation has predicted the absence of these fixed points for $n=2,3$ \cite{Tissier:2000tz}\cite{Tissier:2001uk}. Meanwhile, the situations in lattice Monte-Carlo simulations and experimental results have been equivocal. For a summary of these long-standing discussions in theoretical, numerical and experimental aspects for $n=2,3$, we refer the reader to \cite{delamotte2005fustrated}.

Recently intensive efforts have been made toward non-perturbative understanding of higher dimensional CFTs via the conformal bootstrap program. The output of the program is (within controllable numerical errors) rigorous bounds on scaling dimensions of operators \cite{Rattazzi:2008pe}\cite{Rychkov:2009ij} or operator product expansion coefficients \cite{Caracciolo:2009bx} including various central charges \cite{Poland:2010wg}\cite{Rattazzi:2010gj}\cite{Poland:2011ey}. One eminent feature is the presence of singular behaviors called ``kinks'' in the bounds and the agreement in their positions with previously known interacting CFTs, e.g. Wilson-Fisher fixed points \cite{Rychkov:2009ij}\cite{El-Showk:2012ht}\cite{El-Showk:2013nia}\cite{El-Showk:2014dwa} and $d=3$ $O(N)$ Heisenberg fixed points \cite{Kos:2013tga}. Even without a firm proof, experimental success is convincing enough that the existence of the kink indicates the signal of interacting CFTs sitting there.  

In this Letter we perform the conformal bootstrap program for $O(n)\times O(2)$-symmetric CFTs with a special focus on $n=3$ and $4$ as a natural continuation of our previous work \cite{Nakayama:2014lva}. There we identified all the conjectured fixed points of $O(n)\times O(3)$ symmetric CFTs with sufficiently large $n$ and  proposed the edge of the conformal window for the anti-chiral fixed point.  
Encouraged by these results, we will now tackle the controversies to address the (non)-existence of conjectured CFTs. In both cases, we show a non-perturbative support for the conclusion of \cite{pelissetto2001critical}\cite{Calabrese:2002af}\cite{Calabrese:2004nt}- i.e., the presence of new universality classes that are distinct from Heisenberg ones.

\section{Notations and a quick summary of the $O(n)\times O(3)$-bootstrap results} 
Here we briefly summarize the notations and results of our previous work \cite{Nakayama:2014lva}. We assume the presence of scalar operator $\phi_{a}^\alpha$ (i.e. the elementary field in \eqref{Hamiltonian}) with conformal dimension $\Delta _\phi$ in a bifundamental representation of $O(n)\times O(m)$. The conformal block decomposition of their four-point function has nine independent channels corresponding to the irreducible representations contained in the bifundamental$\times$bifundameltal tensor product, which we label as $
  \mathrm{SS},\ \mathrm{ST}, \ \mathrm{SA}, \ \mathrm{TS}, \cdots ,\ \mathrm{AA}. $
Here $\mathrm{S}, \mathrm{T}, \mathrm{A}$ denote a singlet, a traceless-symmetric tensor, and an anti-symmetric tensor representation of $O(n)$ and $O(m)$ symmetry group, respectively. 

As was first studied in \cite{Poland:2011ey}, we can numerically compute the upper bounds on the dimension of the first operator with definite spin (labelled as $l$) for each sector of the representation (labelled as $R$) which we denote by $\Delta_ c ^{R,l}(\Delta _\phi)$. The upshot of \cite{Nakayama:2014lva} is that for $m=3$ and $n \gg 3$, we can identify all the fixed points as singular behaviors of the output $\Delta _c ^{R,l} (\Delta _\phi)$. In other words, we can ``solve'' these fixed points in the same sense as in \cite{El-Showk:2012ht}\cite{El-Showk:2014dwa}. In addition to the $O(3n)$ Heisenberg fixed point, we found an unstable fixed point  (called ``anti-chiral'' in the literature) in $\Delta_c ^{\mathrm{TA},1}$ and $\Delta _c^{\mathrm{ST},0}$, or a stable fixed point (called ``chiral'' in the literature) in $\Delta _c ^{\mathrm{TS},0}$ but did not find any interesting behaviors in the other sectors. 

Below we compute the bounds for dimensions of various operators in $O(n)\times O(2)$ symmetric CFTs with $n=3$ and $4$, following the scheme of \cite{Kos:2013tga}. We use the hybrid method of Zamolodchikov-type recursion introduced there and extremely convenient  $\rho$-series expansion derived in \cite{Hogervorst:2013kva} to generate partial fractional approximations for conformal blocks. Our sdpa-gmp \cite{sdpa1}\cite{sdpa2} implementation will be identical to the one in \cite{Nakayama:2014lva}. We do not assume intermediate scalar operator dimensions to be greater than 1 unlike in \cite{Kos:2013tga}\cite{El-Showk:2014dwa}, which is obligatory for our purpose because some of the conjectured fixed points have intermediate scalar operators with dimension below 1.

\section{$n=3$ : frustrated spin systems and chiral fixed point}
We first present our results for  $O(3) \times O(2)$-symmetric CFTs, which would describe the anti-ferromagnetic Heisenberg spin systems placed on stacked-triangular lattices. For the LGW model \eqref{Hamiltonian} with this global symmetry, resummed perturbation series studied in \cite{pelissetto2001critical}\cite{Calabrese:2002af}\cite{Calabrese:2004nt}\cite{Calabrese:2004at} predict the presence of two stable fixed points called ``chiral'' ($v>0$) and ``collinear'' ($v<0$) fixed points in addition to Gaussian and Heisenberg ones. On the other hand, functional RG analysis in \cite{Tissier:2000tz} predicts the absence of such fixed points.
\begin{center}
\begin{figure}[h!!]
  \centering
  \includegraphics[width=8cm]{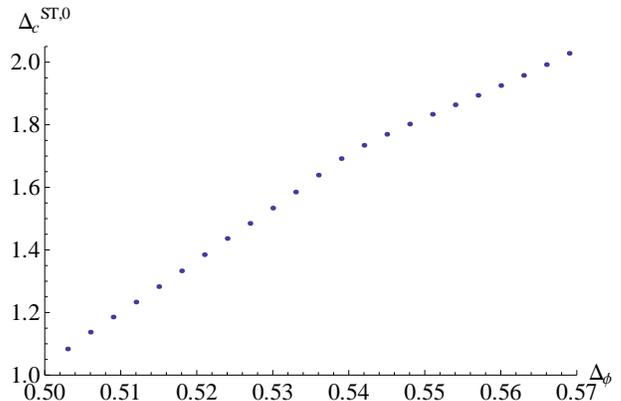}
  \caption{The bound $\Delta_c ^{\mathrm{ST},0}$ for $O(3)\times O(2)$ symmetric CFTs. Here the search space dimension is $55\times 9$, i.e., $k=10$ in the notation of \cite{Kos:2013tga}.}
  \label{fig:1}
\end{figure}
\end{center}
\begin{center}
\begin{figure}[h!!]
  \centering
  \includegraphics[width=8cm]{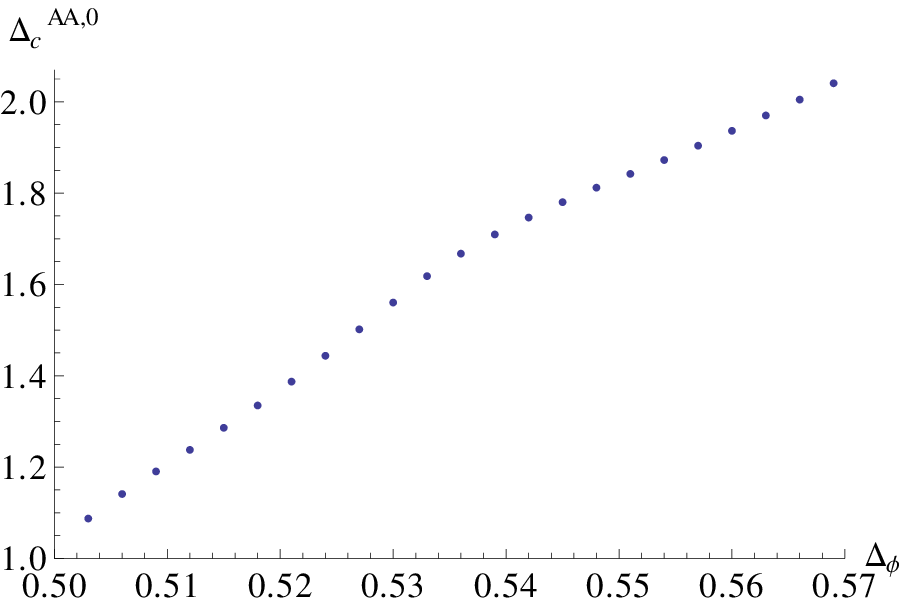}
  \caption{The bound $\Delta_c ^{\mathrm{AA},0}$ for $O(3)\times O(2)$ symmetric CFTs. Here the search space dimension is $36\times 9$, i.e., $k=8$ in the notation of \cite{Kos:2013tga}.}
  \label{fig:2}
\end{figure}
\end{center}

\begin{table}[htbp]
\resizebox{8.8cm}{!}{
\begin{tabular}{|c||c|c|c|c|c|c|}
\hline
 &$\Delta_\phi$ & $\Delta_\mathrm{SS}$ & $\Delta_\mathrm{ST}$ & $\Delta_\mathrm{TS}$ & $\Delta_\mathrm{TT}$ & $\Delta_\mathrm{AA}$ \\ \hline \hline
bootstrap &0.539(3)&1.42(4)&1.69(6)&1.39(3)&1.113(3)&0.89(2)\\ \hline
$\overline{\mathrm{MS}}$ &0.54(2)&1.41(12)&1.79(9)&1.46(8)&1.04(11)&0.75(12)\\ \hline
MZM& 0.55(1)& 1.18 (10) &1.91(5)&1.49(3)&1.01(4)&0.65(13)\\ \hline
\end{tabular}}
\caption{The low-lying spectra read off around the kink in FIG.~\ref{fig:1} and the spectra for the $O(3)\times O(2)$ chiral fixed point from \cite{Calabrese:2004nt} and \cite{Calabrese:2004at}.}
\label{table:1}
\end{table}

\begin{table}[htbp]
\resizebox{8.8cm}{!}{\begin{tabular}{|c||c|c|c|c|c|c|}
\hline
 &$\Delta_\phi$ & $\Delta_\mathrm{SS}$ & $\Delta_\mathrm{ST}$ & $\Delta_\mathrm{TS}$ & $\Delta_\mathrm{TT}$ & $\Delta_\mathrm{AA}$ \\ \hline \hline
bootstrap &0.539(3)&1.39(4)&0.88(3)&1.00(1)&1.234(8)&1.7(1)\\ \hline
$\overline{\mathrm{MS}}$ &0.543(12)& 1.43(20) &0.9(2)&1.0(1)&1.25(5)&1.8(1)\\ \hline
MZM& 0.540(4)& 1.31(10)  &0.95(15)&1.0(2)&1.25(10)&1.75(10)\\ \hline
\end{tabular}}
\caption{The low-lying spectra read off around the kink in FIG.~\ref{fig:2} and the spectra for the $O(3)\times O(2)$ collinear fixed point from \cite{Calabrese:2003ww},\cite{Calabrese:2004nt} and \cite{Calabrese:2004at}.}
\label{table:2}
\end{table}
In FIG.~\ref{fig:1} and \ref{fig:2}, we present $\Delta_c ^{\mathrm{ST},0}$ and $\Delta_c ^{\mathrm{AA},0}$. In both plots we observe kinks around $\Delta_\phi \simeq 0.539$, where we claim that interacting CFTs are located. As was pointed out in \cite{ElShowk:2012hu}, one can read off the  spectra contained in $\phi \times \phi$ OPE. We computed the low-lying scalar spectra in all the intermediate channels which we denote by $\Delta_{R}$ and summarize the values in TABLE \ref{table:1} and \ref{table:2}. To evaluate the systematic errors, we derived the bounds and spectra with lower dimensional search spaces (more specifically, $k=6$ search space to evaluate the error for $k=8$ results, and $k=8$ for $k=10,11$) by estimating the rapidity of convergence. We also took horizontal impreciseness in locating the kink into account.

 While two kinks appeared in the two sectors are close in the values of $\Delta _\phi$, given the spectra of the first operators in various OPE channels, we conclude that they represent two distinct CFTs. In the tables we also quote the values of $\Delta_\phi$ and the intermediate channel spectra for each representation channel by converting the results of \cite{2003PhRvB}\cite{Calabrese:2003ww}\cite{Calabrese:2004nt}\cite{Pelissetto:2013hqa} by $\Delta_\phi = 1/2 + \eta /2$, $\Delta_{\mathrm{SS}} = 3- 1/\nu$ and the TABLE III of \cite{Calabrese:2004at} by $\Delta_{i} = 3 - y_i$ (in their notation, 1=AA, 2=TT, 3 = TS and 4 =ST) and in particular $y_1 = \phi_c/\nu$ with the chiral exponent $\phi_c$ \cite{Kawamura}. Our spectra for the kink in $\Delta_c ^{\mathrm{ST},0}$ agree with the resummed 5-loop expansions in  $\overline{\mathrm{MS}}$ scheme for the chiral fixed point within the systematic error, and those for $\Delta_c ^{\mathrm{AA},0}$ also agree with their results for the collinear fixed point. Although the comparison with the six-loop expansions in MZM scheme is less impressive, it was anticipated in \cite{Calabrese:2004nt} because of the better Borel-summability of $\overline{\mathrm{MS}}$ series.
Our spectra do not show further unstable operators in the SS sector in agreement with the claim that these fixed points are stable. We, however, do not find any indication that the chiral fixed point has a focus point behavior \cite{Calabrese:2002af}.

While the correspondence between kinks and actual CFTs has not been proven yet, we emphasize that our results (as in all the conformal bootstrap studies) are obtained without any reference to the RG analysis based on the Hamiltonian \eqref{Hamiltonian}. We believe that the most natural explanation for such an agreement is the actual existence of these CFTs. We are therefore led to the conjecture: {\it both the chiral and collinear fixed points for $O(3)\times O(2)$ LGW model exist and saturate the bound $\Delta_c ^{\mathrm{ST},0}$ and $\Delta _c ^{\mathrm{AA},0}$, respectively.} The presence of the chiral fixed point with the symmetry breaking pattern $O(3) \times O(2) \to O(2)$
 implies that the phase transition in frustrated spin systems can be continuous.

In the other channels, we do not see any interesting behaviors. We recall that we observed the anti-chiral fixed point in $\Delta_c ^{\mathrm{TA},1}$ for $O(n) \times O(3)$ with $n>7$ \cite{Nakayama:2014lva}. Indeed, we also observe the kink in $\Delta_c ^{\mathrm{TA},1}$  corresponding to the anti-chiral fixed point in $O(10) \times O(2)$ but not in $O(3) \times O(2)$. On the other hand, in $\Delta _c ^{\mathrm{AA},0}$, where we observed a kink for $O(3)\times O(2)$, we do not find any interesting behaviors for $O(10)\times O(2)$. It is plausible that such a ``switching'' behavior of the sectors showing kinks is a reflection of the picture that the anti-chiral fixed point merges into chiral one when $n$ is below a certain critical value $N_{c+}$ and a qualitatively different fixed point (i.e., collinear fixed point) emerges for $n$ below another critical value $N_{c-} \le N_{c+}$.

\section{$n=4$ : QCD and Collinear fixed point}
In \cite{Pisarski:1983ms} it was pointed out that $SU(2)_L \times SU(2)_R$-symmetric LGW models will describe the QCD chiral phase transition which occurred in our early universe, where the bifundamental scalar fields are identified with the meson fields. As was already discussed there, the presence of anomalous $U(1)_A$ symmetry which may be restored by finite-temperature effects could significantly alter the conclusion. The anomaly term makes a certain meson field (including $\eta'$) massive so that the resultant fixed point by tuning the temperature could be only $O(4)$ Heisenberg one (or first order). See \cite{Grahl:2013pba}\cite{Sato:2013tka} for the RG analysis. We point out that some recent numerical as well as theoretical studies suggest the possibility of the effective restoration of $U(1)_A$ at the transition temperature, see e.g.  \cite{Bazavov:2012qja}\cite{Aoki:2012yj}\cite{Cossu:2013uua}\cite{Buchoff:2013nra}\cite{Pelissetto:2013hqa}. If this is the case, the LGW model to be studied is $SU(2)_L\times SU(2)_R \times U(1)_A \simeq O(4)\times O(2)$ as in \eqref{Hamiltonian}, for which the existence of IR-stable fixed point is again controversial. The higher order perturbative study predicts the chiral $(v>0$) and collinear ($v<0$) fixed points. Note that the  {\it collinear} fixed point is the one relevant in the chiral phase transition, where the symmetry breaking pattern is $SU(2)_L \times SU(2)_R \times U(1)_A \to SU(2)_{\mathrm{diag}}$. For the lattice simulations to determine the order of the chiral phase transition with conflicting results, we refer to \cite{Karsch:1993tv}\cite{Iwasaki:1996ya}\cite{AliKhan:2000iz}\cite{D'Elia:2005bv}\cite{Ejiri:2009ac}.

\begin{center}
\begin{figure}[h!!]
  \centering
  \includegraphics[width=8cm]{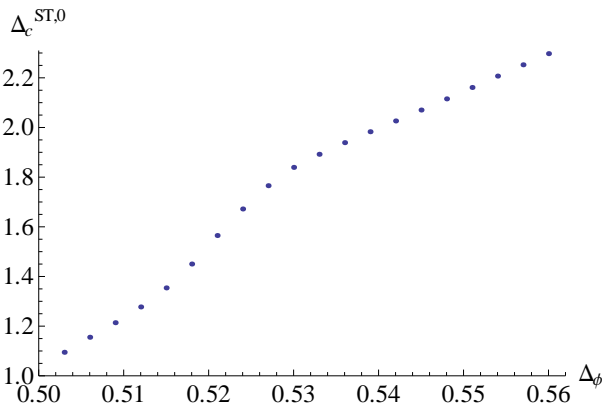}
  \caption{The bound $\Delta_c ^{\mathrm{ST},0}$ for $O(4)\times O(2)$ symmetric CFTs. Here the search space dimension is $36\times 9$, i.e., $k=8$ in the notation of \cite{Kos:2013tga}.}
  \label{fig:3}
\end{figure}
\end{center}
\begin{center}
\begin{figure}[h!!]
  \centering
  \includegraphics[width=8cm]{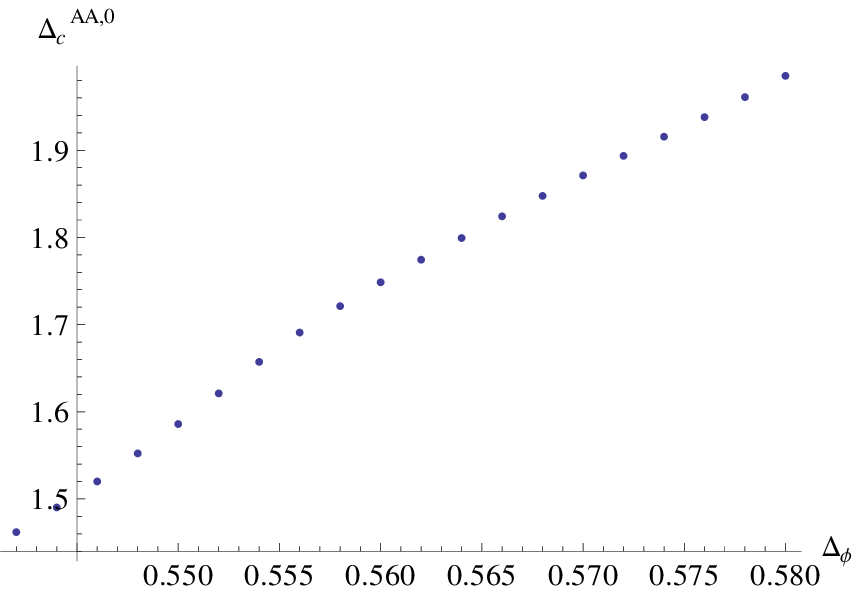}
  \caption{The bound $\Delta_c ^{\mathrm{AA},0}$ for $O(4)\times O(2)$ symmetric CFTs. Here the search space dimension is $66\times 9$, i.e., $k=11$ in the notation of \cite{Kos:2013tga}.}
  \label{fig:4}
\end{figure}
\end{center}
\begin{table}[htbp]
\resizebox{8.8cm}{!}{
\begin{tabular}{|c||c|c|c|c|c|c|}
\hline
 &$\Delta_\phi$ & $\Delta_\mathrm{SS}$ & $\Delta_\mathrm{ST}$ & $\Delta_\mathrm{TS}$ & $\Delta_\mathrm{TT}$ & $\Delta_\mathrm{AA}$ \\ \hline \hline
bootstrap &0.527(3)&1.37(5)&1.76(8)& 1.30(2)&1.084(3)&0.90(1)\\ \hline
$\overline{\mathrm{MS}}$ & 0.536(5) & 1.44(10) &1.83(8)&1.35(3)&1.06(10)&0.83(10)\\ \hline
MZM& 0.533(3) & 1.04(12) &1.94(7) &1.36(5)&0.96(20)&0.71(8)\\ \hline
\end{tabular}}
\caption{The low-lying spectra read off around the kink in FIG.~ \ref{fig:3} and the spectra for the $O(4)\times O(2)$ chiral fixed point from \cite{2003PhRvB},\cite{Calabrese:2004nt} and \cite{Calabrese:2004at}.}
\label{table:3}
\end{table}
\begin{table}[htbp]
\resizebox{8.8cm}{!}{
\begin{tabular}{|c||c|c|c|c|c|c|}
\hline
 &$\Delta_\phi$ & $\Delta_\mathrm{SS}$ & $\Delta_\mathrm{ST}$ & $\Delta_\mathrm{TS}$ & $\Delta_\mathrm{TT}$ & $\Delta_\mathrm{AA}$ \\ \hline \hline
bootstrap &0.556(6)&1.54(6)&0.83(3)&1.044(3)&1.26(2)&1.70(6)\\ \hline
$\overline{\mathrm{MS}}$ &0.56(3)&1.68(17)&1.0(3)&1.10(15)&1.35(10)&1.9(1)\\ \hline
MZM& 0.56(1)& 1.59(14) &0.95(15)&1.25(10)&1.34(5)&1.90(15)\\ \hline
\end{tabular}}
\caption{The low-lying spectra read off around the kink in FIG.~ \ref{fig:4} and the spectra for the $O(4)\times O(2)$ collinear fixed point from \cite{Pelissetto:2013hqa},\cite{Calabrese:2004nt} and \cite{Calabrese:2004at}.}
\label{table:4}
\end{table}

We present our results for $\Delta _c ^{\mathrm{ST},0}$ and $\Delta _c ^{\mathrm{AA},0}$ in FIG \ref{fig:3} and \ref{fig:4}. One technical remark here is that the anti-symmetric tensor representation of $O(4)$ is actually a direct sum of two irreducible representations, but we did not take this into account since the LGW model \eqref{Hamiltonian} relevant for us  has a $\mathbb{Z}_2$-symmetry which permutes $SU(2)_L$ and $SU(2)_R$. 
In TABLE \ref{table:3} and \ref{table:4} we list the spectra at the kink ($\Delta _\phi \simeq 0.527$ for $\Delta_c ^{\mathrm{ST},0}$and $\Delta_\phi \simeq 0.556$ for $\Delta_c ^{\mathrm{AA},0}$). From this comparison we find it reasonable to regard the kink in $\Delta _c^{\mathrm{ST},0}$ as the chiral fixed point and that in $\Delta _c ^{\mathrm{AA},0}$ as the collinear fixed point. Hence our non-perturbative results in agreement with the higher-loop analysis in RG provides a strong support for the existence of the chiral as well as collinear fixed point, and the latter, in particular, suggests the possibility of continuous chiral phase transition in QCD once the $U(1)_A$ is effectively restored.


\section{Discussions} 
In this Letter we have performed the conformal bootstrap program for $O(n)\times O(2)$-symmetric CFTs with $n=3$ and $4$. As in $n \gg m =3$ case carried out in \cite{Nakayama:2014lva}, we have observed singular behaviors in the bounds of the dimension of operators in different sectors. Although our identification for them to be actual CFTs is still phenomenological based on the past experiences, their agreement with earlier higher-loop perturbative results is a striking  evidence for the validity of both methods. We, therefore, believe that the results in \cite{pelissetto2001critical}\cite{Calabrese:2003ww}\cite{Calabrese:2004nt} together with ours are robust enough to conclude that the systems inside attractor regions in their RG flows should exhibit continuous phase transitions with the critical exponents that we most precisely predict from the output of the conformal bootstrap program.

An obvious future direction is to study the bootstrap constraints for $O(2)\times O(2)$-symmetric CFTs, from which we expect to gain some insights about the controversial phase transition of $XY$-frustrated spin systems. Due to the emergence of $\mathbb{Z}_2$-symmetry which permutes two $O(2)$-representation, however, we anticipate the bootstrap equations are degenerated and less information can be extracted. 

We end with several challenges to the conformal bootstrap program. Firstly, we had to resort to the earlier RG results to determine the symmetry breaking pattern. To be fully self-contained, it is imperative to know the signs of some OPE coefficients, but this is impossible from the bootstrap of the single correlator alone since all the OPE coefficients appear as the squares of them. The mixed-correlator bootstrap \cite{Kos:2014bka} with the energy-momentum tensor will fix the sign issue. However, the independent structures in the four-point function of the energy-momentum tensors are numerous \cite{Dymarsky:2013wla}, and the computational task would be even more intensive. Secondly we are urged to understand the meaning of kinks, for which the recently observed ``spectrum jumping'' behavior might be helpful \cite{El-Showk:2014dwa}. It has also been observed in \cite{Kos:2014bka} that the simultaneous consideration of several correlators and assumptions on the number of relevant operators in the spectrum is so restrictive that it singles out the region around the kink (where the $d=3$ Ising model is supposed to live) as an isolated ``island'' in the space of entire $\mathbb{Z}_2$-symmetric CFTs. Our conclusion would become more convincing if we observed similar phenomena also in our models. All in all, it is quite plausible that deeper knowledge of the conformal bootstrap program will revolutionalize our understanding of the world. Then the resolutions of the controversies in QCD and frustrated spin systems would be just a small example.

\section*{Acknowledgements}
We would like to thank P.~Calabrese, B.~Delamotte,  D.~Mouhanna, S.~Rychkov, D.~Simmons-Duffin and E.~Vicari for correspondence and discussions. Y.~N. also thanks Y.~Iwasaki, S.~Rey and N.~Yamada for sharing thoughts.
This work is supported by the World Premier International Research Center Initiative (WPI Initiative), MEXT. T.O. is supported by JSPS Research Fellowships for Young Scientists and the Program for Leading Graduate Schools, MEXT.
\providecommand{\href}[2]{#2}\begingroup\raggedright\endgroup

\end{document}